\documentclass[aps,prb,twocolumn,superscriptaddress,nofootinbib,floatfix,longbibliography]{revtex4-2}

\usepackage{amsmath,amssymb,bm,mathtools}
\usepackage{graphicx}
\usepackage{xcolor}
\usepackage{placeins}
\usepackage[colorlinks=true,linkcolor=blue,citecolor=blue,urlcolor=blue]{hyperref}
\allowdisplaybreaks

\newcommand{\Liou}{\mathcal{L}}
\newcommand{\dd}{\mathrm{d}}
\newcommand{\re}{\mathrm{Re}}

\newcommand{\sh}{\mathrm{sh}}
\newcommand{\defect}{\mathrm{def}}
\newcommand{\crit}{\mathrm{crit}}
\newcommand{\safeincludegraphics}[2][]{%
  \IfFileExists{#2}{\includegraphics[#1]{#2}}{%
    \fbox{\parbox[c][0.16\textheight][c]{0.92\linewidth}{\centering Missing figure\\\texttt{\detokenize{#2}}}}%
  }%
}

\begin{document}

\title{Microscopic resonant-shell mechanism for slow Liouvillian sectors in an open correlated lattice}
\author{X. Z. Zhang}
\email{zhangxz@tjnu.edu.cn}
\affiliation{College of Physics and Materials Science, Tianjin Normal University, Tianjin 300387, China}

\begin{abstract}
We develop a microscopic theory for how slow Liouvillian sectors are selected in an open correlated lattice. The starting point is not a postulated non-Hermitian band, but a local interacting resonance between an on-site doublon and a branch-resolved nearest-neighbor bond. This resonance defines a composite shell orbital whose doublon weight controls reservoir visibility and whose mixed doublon-bond character controls shell mobility. Projecting the microscopic hopping onto the selected shell yields a branch-selective dimerized channel. In the dilute regime, a boundary doublon-loss channel yields an exponentially slow edge-memory pole through a Zeno-type return. At the shell-critical point, the edge pole is replaced by a near-zero standing-wave doublet with an algebraic coherent spacing. At finite shell filling, the same local shell becomes density dressed. A number-conserving phase-locking jump removes a bright mismatch sector, leaving defects as the asymptotic slow variables and producing a diffusive finite-size gap. We derive the local shell, the projected branch topology, the edge-memory law, the shell-critical doublet, the density-dressed shell Hamiltonian, and the defect generator within one Schur-projection framework. The resulting mechanism identifies the reservoir-engineered fast block as the selector of the observable slow sector, while the microscopic parent shell remains fixed.
\end{abstract}

\maketitle

\section{Introduction}

Engineered dissipation has become a constructive ingredient of quantum many-body physics. 
Rather than merely degrading coherence, a reservoir can reshape the long-time Hilbert-space structure by selecting which modes, correlations, and composite objects remain visible after fast degrees of freedom have decayed. This selection problem is encoded in the low-lying Liouvillian spectrum and is directly tied to metastability, boundary memory, anomalous relaxation, and the preparation of nonequilibrium steady states~\cite{Diehl2008,Verstraete,Diehl2011,Daley2014}. 
It becomes especially nontrivial in interacting open lattices, where the slow object is often not a bare single-particle mode, but a correlation-dressed excitation whose visibility depends simultaneously on the microscopic Hamiltonian and on the reservoir algebra.

A common route to dissipative topology and slow relaxation starts from a target dark state, an effective non-Hermitian band, or a reduced Liouvillian matrix~\cite{Budich,IeminiPRB,Macieszczak,MoriShirai,Nosov,Bardyn2012,LangBuchler2015,Goldstein2019,Beck2021,Liu2021,Talkington2022,Xu2023,Yang2023,Tonielli2020,Shen2014,RudnerLevitov2009,Lee2016,Gong2018,YaoWang2018,Kunst2018}. 
These descriptions are powerful once the relevant effective degrees of freedom are known. 
They do not, by themselves, explain why a particular microscopic object is retained as the slow degree of freedom after Hamiltonian detuning, interaction constraints, and dissipative fast sectors are eliminated. In an interacting lattice this microscopic selection problem is central, because the long-lived object may be neither a bare particle nor a bare doublon, but a local resonant composite that carries both coherent mobility and reservoir visibility.

In this work we develop a shell-first mechanism for slow Liouvillian-sector selection.  The microscopic Hamiltonian first produces a local resonance between an on-site doublon and a branch-resolved nearest-neighbor bond.  This resonance defines a hybrid shell orbital with two independent physical roles: its doublon weight controls reservoir visibility, while its mixed doublon-bond character controls shell mobility.  Projection of microscopic hopping turns the same shell into a branch-selective dimerized channel in the dilute regime and into a density-dressed shell Hamiltonian at finite filling.  The engineered jump algebra then determines which fast block is removed by the Liouvillian Schur return.

\begin{figure}[!t]
\centering
\safeincludegraphics[width=0.98\columnwidth]{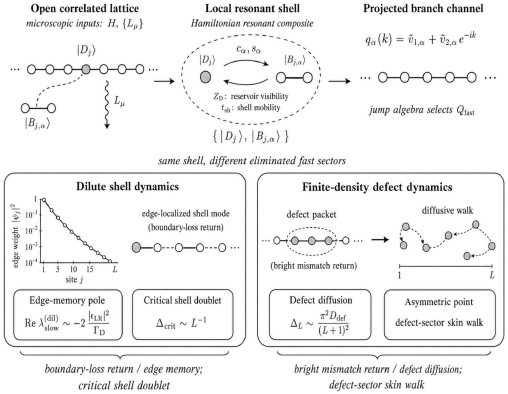}
\caption{Shell-first organization of the slow sectors.  The microscopic Hamiltonian and engineered jump algebra first select a local resonant shell $\{|D_j\rangle,|B_{j,\alpha}\rangle\}$.  Its composition fixes the reservoir visibility $Z_{D,\alpha}$ and the shell mobility $t_{\sh}$.  Projection of microscopic hopping gives the branch channel $q_\alpha(k)$, while different eliminated fast sectors produce the dilute boundary-loss return and the finite-density bright-mismatch return.  The figure represents one parent shell acted on by different fast-sector eliminations, rather than a collection of unrelated effective models.}
\label{fig:fig1}
\end{figure}

The same microscopic shell gives three different slow-sector realizations.  A boundary doublon-loss channel produces an exponentially slow edge-memory pole.  At the projected shell-critical point, the edge pole is replaced by a near-zero standing-wave doublet with an algebraic coherent spacing.  At finite shell filling, a number-conserving phase-locking jump removes a bright mismatch block and leaves defects as the asymptotic slow degrees of freedom, giving a diffusive finite-size gap.  Figure~\ref{fig:fig1} summarizes this organizational logic: a microscopic shell is selected first, and the observable slow sector is then chosen by the fast block eliminated by the reservoir-engineered protocol.

The remainder of this paper is organized as follows.  Section~\ref{sec:model} introduces the microscopic open lattice and derives the local resonant shell.  Section~\ref{sec:branch} projects the microscopic hopping onto the shell and obtains the branch-selective dimerized channel.  Section~\ref{sec:dilute} derives the dilute edge-memory law.  Section~\ref{sec:critical} analyzes the shell-critical coherent doublet and the boundary-loss-induced two-mode non-Hermitian problem.  Section~\ref{sec:density} develops the density-dressed shell Hamiltonian and the microscopic route to phase locking.  Section~\ref{sec:defect} eliminates the bright mismatch block and obtains the defect diffusion law, including the unreduced shell-many-body bridge.  Section~\ref{sec:skin} discusses the asymmetric defect-sector skin walk.  Section~\ref{sec:discussion} summarizes the mechanism, its regime of validity, and directly observable signatures.  The appendices provide complementary diagnostics: Appendix~\ref{app:defect-msd} gives a reduced-shell time-domain check of defect diffusion, Appendix~\ref{app:rate-benchmark} benchmarks the locked-plus-bright Schur return, and Appendix~\ref{app:skin-benchmark} gives a minimal asymmetric-skin benchmark.

\section{Microscopic model and local resonant shell}
\label{sec:model}

We consider a one-dimensional interacting lattice governed by a Markovian master equation \cite{Gorini1976,Lindblad},
\begin{equation}
\begin{aligned}
\dot\rho
&=\Liou[\rho]
=-i[H,\rho]+\sum_\mu \mathcal D[L_\mu]\rho,\\
\mathcal D[L]\rho
&=L\rho L^\dagger-\frac12\{L^\dagger L,\rho\}.
\end{aligned}
\label{eq:master}
\end{equation}
The Hamiltonian is written explicitly as
\begin{equation}
H=H_t+H_U+H_V+H_\delta+H_h,
\label{eq:H_micro_decomp}
\end{equation}
with
\begin{equation}
\begin{aligned}
H_t&=-\sum_{j,\sigma}\kappa_\sigma
\left(c_{j+1,\sigma}^\dagger c_{j,\sigma}+\mathrm{H.c.}\right),\\
H_U&=U\sum_j n_{j\uparrow}n_{j\downarrow},\qquad
H_V=V\sum_j n_j n_{j+1},\\
H_\delta&=-i\delta\sum_j
\left(s_j^-s_{j+1}^+ -s_j^+s_{j+1}^-\right),\\
H_h&=h\sum_j(-1)^j s_j^z.
\end{aligned}
\label{eq:H_explicit}
\end{equation}
Here $c_{j\sigma}$ annihilates a fermion with spin $\sigma=\uparrow,\downarrow$ on site $j$, $n_{j\sigma}=c_{j\sigma}^\dagger c_{j\sigma}$, $n_j=n_{j\uparrow}+n_{j\downarrow}$, and $s_j^a$ are spin operators.  The hopping amplitudes $\kappa_\uparrow$ and $\kappa_\downarrow$ need not be equal.  The term $H_\delta$ mixes opposite-spin nearest-neighbor pair states on a bond, while $H_h$ splits them in a staggered way.  The engineered jumps $L_\mu$ are introduced below after the shell has been identified, because their physical action is best described after projection.

The roles of the microscopic terms are already separated at this stage.  The interaction terms $H_U$ and $H_V$ set the bare doublon and nearest-neighbor pair energies.  The staggered field $H_h$ and the chiral bond term $H_\delta$ split the nearest-neighbor pair into two branch orbitals.  The spin-dependent hopping $H_t$ is not merely a kinetic term added after the shell has formed; after projection it determines the branch-dependent dimerization and the mobility scale of the retained shell.  Thus all ingredients of the later effective theory are present at the microscopic level: resonance, branch selection, reservoir visibility, and shell motion.

The slow sector is selected in the two-particle problem before any SSH reduction is invoked.  The local doublon and nearest-neighbor opposite-spin pair states are
\begin{equation}
|D_j\rangle=c_{j\uparrow}^\dagger c_{j\downarrow}^\dagger|0\rangle,
\label{eq:D_state}
\end{equation}
\begin{equation}
|X_j\rangle=c_{j\uparrow}^\dagger c_{j+1,\downarrow}^\dagger|0\rangle,
\qquad
|Y_j\rangle=c_{j\downarrow}^\dagger c_{j+1,\uparrow}^\dagger|0\rangle.
\label{eq:XY_states}
\end{equation}
The doublon carries interaction energy $U$.  On bond $(j,j+1)$ the nearest-neighbor pair subspace is governed by
\begin{equation}
H_{B,j}
=
V\mathbb{I}_2+
\begin{pmatrix}
h_j&i\delta\\
-i\delta&-h_j
\end{pmatrix},
\qquad h_j=h(-1)^j,
\label{eq:bond_block_main}
\end{equation}
where the basis is $\{|X_j\rangle,|Y_j\rangle\}$.
The entries in Eq.~\eqref{eq:bond_block_main} follow directly from the two local terms that split the bond.  The staggered Zeeman field contributes
\begin{equation}
H_{h,j}=h(-1)^j s_j^z+h(-1)^{j+1}s_{j+1}^z
=h_j(s_j^z-s_{j+1}^z),
\end{equation}
which gives the diagonal matrix $\mathrm{diag}(h_j,-h_j)$ in the basis $\{|X_j\rangle,|Y_j\rangle\}$.  The chiral bond term may be written as
\begin{equation}
H_{\delta,j}=-i\delta\left(s_j^-s_{j+1}^+-s_j^+s_{j+1}^-\right),
\end{equation}
which gives the off-diagonal matrix
\begin{equation}
H_{\delta,j}=\begin{pmatrix}0&i\delta\\-i\delta&0\end{pmatrix}_{\{|X_j\rangle,|Y_j\rangle\}} .
\end{equation}
This explicit local diagonalization is the first few-body step of the shell-first construction.  It turns the single nearest-neighbor bond sector into two branch-resolved resonance conditions rather than one undifferentiated doublon-bond crossing.
Its eigenvalues are
\begin{equation}
E_{B,\alpha}=V+\alpha\Omega,
\qquad
\Omega=\sqrt{h^2+\delta^2},
\qquad
\alpha=\pm.
\label{eq:branch_energy_main}
\end{equation}
Thus the original doublon-bond resonance $U\simeq V$ is resolved into two branch conditions $U\simeq V\pm\Omega$.

It is convenient to introduce an angle $\beta_j$ through $\cos\beta_j=h_j/\Omega$ and $\sin\beta_j=\delta/\Omega$.  One normalized choice of branch eigenvectors is
\begin{equation}
\begin{aligned}
|B_{j,+}\rangle&=\cos\frac{\beta_j}{2}|X_j\rangle
-i\sin\frac{\beta_j}{2}|Y_j\rangle,\\
|B_{j,-}\rangle&=-i\sin\frac{\beta_j}{2}|X_j\rangle
+\cos\frac{\beta_j}{2}|Y_j\rangle.
\end{aligned}
\label{eq:branch_vectors_main}
\end{equation}
For a chosen branch $\alpha$, the local doublon-branch block is
\begin{equation}
h^{\rm loc}_{j,\alpha}=
\begin{pmatrix}
U&g_\alpha\\
g_\alpha^*&V+\alpha\Omega
\end{pmatrix},
\qquad
\Delta_\alpha=U-(V+\alpha\Omega),
\label{eq:local_block_main}
\end{equation}
where $g_\alpha$ is the projected microscopic conversion matrix element.  The lower shell eigenenergy is
\begin{equation}
\varepsilon_{\sh,\alpha}=\frac{U+V+\alpha\Omega}{2}
-\frac12\sqrt{\Delta_\alpha^2+4|g_\alpha|^2}.
\label{eq:shell_energy_main}
\end{equation}
The retained shell orbital is
\begin{equation}
p_j^\dagger=\cos\vartheta_\alpha D_j^\dagger
+e^{i\phi_\alpha}\sin\vartheta_\alpha B_{j,\alpha}^\dagger,
\label{eq:shell_orbital_main}
\end{equation}
with
\begin{equation}
\tan 2\vartheta_\alpha=\frac{2|g_\alpha|}{\Delta_\alpha},
\qquad e^{i\phi_\alpha}=g_\alpha/|g_\alpha|.
\label{eq:shell_angle_main}
\end{equation}
The shell weights are
\begin{equation}
Z_{D,\alpha}=|\langle D_j|p_j\rangle|^2=\cos^2\vartheta_\alpha,
\qquad
Z_{B,\alpha}=\sin^2\vartheta_\alpha.
\label{eq:shell_weights_main}
\end{equation}
This hybrid composition is the first key microscopic fact.  The doublon weight $Z_{D,\alpha}$ controls how strongly a doublon-selective reservoir sees the shell, while the mixed doublon-bond character controls how the shell moves under the single-particle hopping $H_t$.  Consequently the same resonance angle reappears in the dilute edge-memory scale and in the finite-density diffusion scale.

The local diagonalization therefore produces three quantities that will reappear throughout the paper: the shell energy $\varepsilon_{\sh,\alpha}$, the reservoir visibility $Z_{D,\alpha}$, and the shell mobility generated after projecting $H_t$.  The first identifies the resonant manifold, the second determines how the reservoir sees the shell, and the third determines how the shell moves in both dilute and finite-density settings.  This is why the local shell is more than a convenient basis choice: it is the microscopic object on which both coherent and dissipative reductions act.

\section{Projected branch channel and shell topology}
\label{sec:branch}

The shell orbital is not yet an SSH chain.  Dimerization appears only after projecting $H_t$ onto the selected shell branch.  Because $h_j=(-1)^j h$ alternates, the bond eigenvectors on even and odd bonds differ, and the same microscopic hopping produces two inequivalent projected amplitudes.  Writing
\begin{equation}
\kappa_\pm=\frac{\kappa_\uparrow\pm\kappa_\downarrow}{2},
\label{eq:kappapm_main}
\end{equation}
the projected bond norm is
\begin{equation}
\tilde v_{j,\alpha}^{\;2}
=\kappa_+^2+\kappa_-^2-\frac{2\alpha h_j}{\Omega}\kappa_+\kappa_-.
\label{eq:vj_norm_main}
\end{equation}

\begin{figure*}[!t]
\centering
\safeincludegraphics[width=0.82\textwidth]{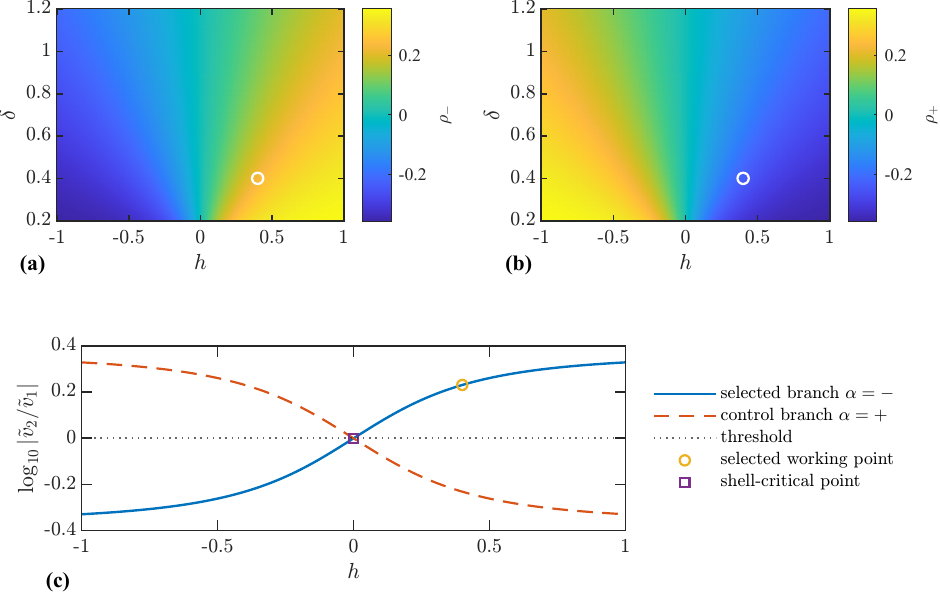}
\caption{Branch-selective dimerization maps used to choose the dilute and shell-critical working points.  Panels (a) and (b) show $\rho_- = \log_{10}|\tilde v_{2,-}/\tilde v_{1,-}|$ and $\rho_+ = \log_{10}|\tilde v_{2,+}/\tilde v_{1,+}|$ over the $(h,\delta)$ plane for $(\kappa_\uparrow,\kappa_\downarrow)=(1.0,0.4)$.  Panel (c) shows the fixed-$\delta$ cut at $\delta=0.40$.  The filled circle marks the selected topological working point $(h,\delta)=(0.40,0.40)$ and the open square marks the shell-critical point $(h,\delta)=(0,0.40)$.  The shell-topological threshold is $|\tilde v_2/\tilde v_1|=1$.}
\label{fig:branchmaps}
\end{figure*}

The rewriting from Eq.~\eqref{eq:vj_norm_main} to Eq.~\eqref{eq:v12_main} uses
\begin{equation}
\kappa_+^2+\kappa_-^2=\frac{\kappa_\uparrow^2+\kappa_\downarrow^2}{2},
\qquad
2\kappa_+\kappa_-=\frac{\kappa_\uparrow^2-\kappa_\downarrow^2}{2}.
\end{equation}
This step is important because it shows explicitly that unequal microscopic spin hoppings are enough to generate branch-dependent dimerization after projection.  No alternating bare hopping has to be inserted by hand.
The two alternating couplings can be written directly in terms of the microscopic spin-dependent hoppings,
\begin{equation}
\begin{aligned}
\tilde v_{1,\alpha}^{\;2}
&=\frac{1+\alpha h/\Omega}{2}\kappa_\uparrow^2
+\frac{1-\alpha h/\Omega}{2}\kappa_\downarrow^2,\\
\tilde v_{2,\alpha}^{\;2}
&=\frac{1-\alpha h/\Omega}{2}\kappa_\uparrow^2
+\frac{1+\alpha h/\Omega}{2}\kappa_\downarrow^2.
\end{aligned}
\label{eq:v12_main}
\end{equation}
After fixing a real gauge for the projected shell basis, we take $\tilde v_{1,\alpha}$ and $\tilde v_{2,\alpha}$ as the positive square roots of the corresponding projected norms; possible overall signs are absorbed into the sublattice basis.  The projected channel now has the form of an SSH problem, but the dimerization has not been assumed.  It is generated by the mismatch between branch eigenvectors on neighboring bonds together with the spin dependence of the microscopic hopping.  This is why the two branches can have different topological indices under the same microscopic parameters.
The corresponding chiral Bloch Hamiltonian is
\begin{equation}
h_\alpha(k)=
\begin{pmatrix}
0&q_\alpha(k)\\
q_\alpha^*(k)&0
\end{pmatrix},
\qquad
q_\alpha(k)=\tilde v_{1,\alpha}+\tilde v_{2,\alpha}e^{-ik}.
\label{eq:branch_bloch_main}
\end{equation}
The shell-topological index is the winding
\begin{equation}
W_\alpha=\frac{1}{2\pi i}\int_{-\pi}^{\pi}\dd k\,\partial_k\ln q_\alpha(k),
\label{eq:winding_main}
\end{equation}
which equals one for $|\tilde v_{2,\alpha}|>|\tilde v_{1,\alpha}|$ and zero for $|\tilde v_{2,\alpha}|<|\tilde v_{1,\alpha}|$.  Figure~\ref{fig:branchmaps} shows the corresponding branch-selective dimerization maps.  The two branches need not share the same topological sector at the same microscopic parameters.  The filled marker selects the dilute topological working point, while the open marker gives the projected shell-critical point at which $|\tilde v_{1,\alpha}|=|\tilde v_{2,\alpha}|$. 

\section{Dilute branch and edge-memory pole}
\label{sec:dilute}

A boundary doublon-selective loss projects onto the retained shell as a
loss channel from the boundary shell orbital to the vacuum.  More explicitly,
let
\[
P_{0+\sh}=|0\rangle\langle 0|+\sum_m |p_m\rangle\langle p_m|
\]
be the projector onto the vacuum and the retained one-shell manifold.  Then
\begin{equation}
P_{0+\sh}L^{(d)}P_{0+\sh}
\simeq
\sqrt{\gamma_d}\,\langle D_1|p_1\rangle\, p_1,
\qquad
|\langle D_1|p_1\rangle|^2=Z_{D,\alpha}.
\label{eq:boundary_loss_projection}
\end{equation}
After absorbing the local phase of $\langle D_1|p_1\rangle$ into the
definition of the projected jump, this channel can be written as
\begin{equation}
L_{\sh}^{(d)}
\simeq
\sqrt{\gamma_d Z_{D,\alpha}}\,p_1 .
\label{eq:projected_boundary_loss}
\end{equation}
This is the first place where the local resonance reappears in a
Liouvillian decay channel: the reservoir does not see an abstract SSH edge
state, but only the doublon component of the shell.  A microscopic
auxiliary-channel implementation of this boundary doublon-selective loss is
described in Sec.~\ref{subsec:boundary-loss-implementation}. For open boundary conditions and $|\tilde v_{1,\alpha}|<|\tilde v_{2,\alpha}|$, the projected branch channel has left and right boundary states.  Here $|A_m\rangle$ and $|B_m\rangle$ denote the two sublattice shell states in unit cell $m$ of the projected branch channel.
\begin{equation}
|L_0\rangle={\cal N}\sum_{m=1}^{M}r_\alpha^{m-1}|A_m\rangle,
\qquad
|R_0\rangle={\cal N}\sum_{m=1}^{M}r_\alpha^{M-m}|B_m\rangle,
\end{equation}
where
\begin{equation}
r_\alpha=-\frac{\tilde v_{1,\alpha}}{\tilde v_{2,\alpha}},
\qquad
{\cal N}^2=\frac{1-r_\alpha^2}{1-r_\alpha^{2M}}.
\end{equation}
Their finite-size communication amplitude is
\begin{equation}
\varepsilon_{LR}=\tilde v_{1,\alpha}
\frac{1-r_\alpha^2}{1-r_\alpha^{2M}}r_\alpha^{M-1}.
\end{equation}
With $\Gamma_D=\gamma_d Z_{D,\alpha}$, the reduced non-Hermitian edge block is
\begin{equation}
H_{\rm edge}=\begin{pmatrix}
-i\Gamma_D/2&\varepsilon_{LR}\\
\varepsilon_{LR}&0
\end{pmatrix}.
\label{eq:Hedge_main}
\end{equation}
Let $E$ be an eigenvalue of this block.  The secular equation is
\begin{equation}
E\left(E+\frac{i\Gamma_D}{2}\right)-|\varepsilon_{LR}|^2=0.
\end{equation}
In the Zeno window $|\varepsilon_{LR}|\ll \Gamma_D$, the fast root is $E_{\rm fast}\simeq -i\Gamma_D/2$, while the slow root satisfies $E_{\rm slow}(i\Gamma_D/2)\simeq |\varepsilon_{LR}|^2$.  The slow Liouvillian decay rate is therefore
\begin{equation}
\re\lambda_{\rm slow}^{({\rm dil})}
\simeq -\frac{2|\varepsilon_{LR}|^2}{\Gamma_D}.
\label{eq:edge_memory_main}
\end{equation}
Strong boundary loss therefore protects the asymptotic memory rather than simply destroying it: the slow decay is a virtual return suppressed by the fast lossy sector.  Topology supplies the exponentially weak communication between edges, while the resonance supplies the doublon weight through which the reservoir can act.  This explicit boundary-state reduction shows why the memory scale is simultaneously topological, through $\varepsilon_{LR}$, and microscopic, through $Z_{D,\alpha}$.

\section{Shell-critical coherent doublet}
\label{sec:critical}

At $|\tilde v_{1,\alpha}|=|\tilde v_{2,\alpha}|\equiv t_c$, the projected branch becomes an open uniform shell chain of length $2M$.  Its coherent Hamiltonian is
\begin{equation}
H_{\crit}^{({\rm coh})}=-t_c\sum_{\ell=1}^{2M-1}
\left(|\ell+1\rangle\langle \ell|+|\ell\rangle\langle \ell+1|\right).
\label{eq:Hcrit_main}
\end{equation}
The standing-wave eigenmodes and eigenvalues are
\begin{equation}
\begin{aligned}
\phi_n(\ell)&=\sqrt{\frac{2}{2M+1}}\sin\frac{n\pi\ell}{2M+1},\\
E_n&=-2t_c\cos\frac{n\pi}{2M+1}.
\end{aligned}
\label{eq:standing_main}
\end{equation}
The pair closest to zero is $n=M,M+1$, and its spacing is
\begin{equation}
\Delta_{\crit}(M)=|E_{M+1}-E_M|
=4t_c\sin\frac{\pi}{4M+2}
\simeq \frac{2\pi t_c}{2M+1}.
\label{eq:Dcrit_main}
\end{equation}
This $L^{-1}$ scale is the shell-critical coherent spacing, not the absolute Liouvillian gap of the full boundary-damped critical ladder.  It is the spacing selected by the near-zero shell doublet, by the initial state, and by the observable.

\begin{figure}[!t]
\centering
\safeincludegraphics[width=0.5\textwidth]{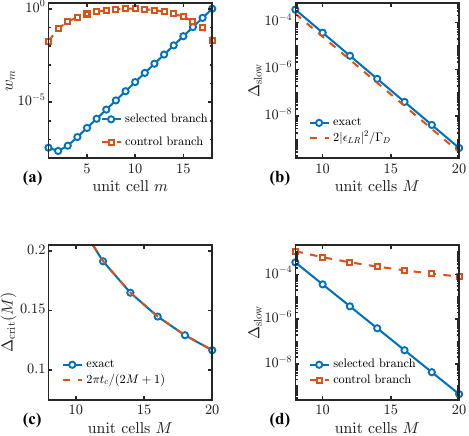}
\caption{Dilute and shell-critical diagnostics.  The selected branch is $\alpha=-$ at $(\kappa_\uparrow,\kappa_\downarrow,h,\delta,\gamma_d)=(1.0,0.4,0.40,0.40,0.40)$, and the control branch is $\alpha=+$.  Panel (a) shows the normalized slow-mode weight $w_m$ for $M=18$.  Panel (b) compares the selected reduced slow gap $\Delta_{\rm slow}$ with the edge-memory estimate $2|\varepsilon_{LR}|^2/\Gamma_D$.  Panel (c) shows the shell-critical spacing with $2\pi t_c/(2M+1)$, and panel (d) contrasts selected and control slow gaps.}
\label{fig:fig2}
\end{figure}

The projected boundary loss on the near-zero doublet produces a useful non-Hermitian two-level reduction.  In the basis of the two near-zero standing waves, the effective Hamiltonian may be written as
\begin{equation}
H_{\rm pair}^{\rm eff}=\frac{\Delta_{\crit}(M)}{2}\sigma_z
-\frac{i\Gamma_{\rm pair}}{2}\left(\mathbb{I}+\sigma_x\right),
\label{eq:Hpair_main}
\end{equation}
where $\Gamma_{\rm pair}\propto \Gamma_D/(2M+1)$ follows from the boundary weight of the near-zero standing waves.  Its eigenvalues are
\begin{equation}
\lambda_{\pm}^{\rm pair}=-\frac{i\Gamma_{\rm pair}}{2}
\pm \frac12\sqrt{\Delta_{\crit}(M)^2-\Gamma_{\rm pair}^2}.
\label{eq:lpair_main}
\end{equation}
The shell-critical doublet therefore crosses from split to overdamped dynamics when the boundary-induced width matches the coherent spacing.  Because both $\Gamma_{\rm pair}$ and $\Delta_{\crit}$ scale as $L^{-1}$, this competition is scale matched within the selected doublet.  
The distinction from a global Liouvillian gap can be checked directly from the same standing-wave basis.  Boundary loss at site $1$ gives widths proportional to $|\phi_n(1)|^2$, with
\begin{equation}
|\phi_n(1)|^2=\frac{2}{2M+1}\sin^2\frac{n\pi}{2M+1}.
\end{equation}
For the smallest-wave-vector mode, $n=1$, this scales as $(2M+1)^{-3}$, so the full critical ladder can contain boundary-induced decay widths smaller than the $L^{-1}$ width of the near-zero doublet.  This is why $\Delta_{\crit}$ is used here as a shell-critical coherent spacing rather than as a separately established global Liouvillian gap.
Figure~\ref{fig:fig2} displays the dilute and critical diagnostics.

\section{Finite-density shell Hamiltonian and engineered phase locking}
\label{sec:density}

At finite shell filling the shell is not replaced by a new model; it is dressed.  The local resonant orbital remains the same microscopic object, but the background occupation changes its transfer amplitude, residual interaction, and the identity of its long-lived excitations.  We therefore keep the same shell orbital and generate the finite-density Hamiltonian by a Hilbert-space Schur map,
\begin{equation}
\begin{aligned}
H_{\sh}^{(N)}={}&
\mu_{\sh}(\nu)\sum_j n_j
-t_{\sh}(\nu)\sum_j
\left(e^{iQ}p_{j+1}^{\dagger}p_j+\mathrm{H.c.}\right)\\
&+U_{\sh}(\nu)\sum_j n_j n_{j+1}+\cdots.
\end{aligned}
\label{eq:HshN_main}
\end{equation}
Here $n_j=p_j^\dagger p_j$, and $Q$ denotes the relative phase selected by the locked finite-density shell configuration.  The finite-density coefficients are obtained by applying the same shell projector before eliminating shell-breaking configurations. The calculation has three layers. The local resonance fixes the shell orbital; direct projection gives the zeroth-order coefficients $\mu_{\sh}^{(0)}$, $t_{\sh}^{(0)}$, and $U_{\sh}^{(0)}$; virtual excursions into $Q_N$ then dress them into $\mu_{\sh}(\nu)$, $t_{\sh}(\nu)$, and $U_{\sh}(\nu)$. This separation is useful because it keeps the microscopic origin of each effective coefficient visible.

\subsection{Zeroth-order projected coefficients}
At zeroth order in inter-shell virtual excursions, the shell chemical potential is simply the lower eigenvalue of the local resonant block,
\begin{equation}
\mu_{\sh}^{(0)}=\varepsilon_{\sh,\alpha}=\frac{U+V+\alpha\Omega}{2}-\frac12\sqrt{\Delta_\alpha^2+4|g_\alpha|^2}.
\end{equation}
This formula already displays the dependence on the microscopic Hamiltonian parameters: $U$ enters through the doublon energy, $V\pm\Omega$ through the branch-resolved bond energies, and $g_\alpha$ through the local doublon--bond conversion. In other words, even before any many-body dressing is introduced, the shell chemical potential is already a microscopic resonance quantity rather than a free effective parameter.

Writing
\begin{equation}
p_j^\dagger=c_\alpha D_j^\dagger+s_\alpha e^{i\phi_\alpha}B_{j,\alpha}^\dagger,
\qquad
c_\alpha=\cos\vartheta_\alpha,\quad s_\alpha=\sin\vartheta_\alpha,
\end{equation}
the coherent shell transfer is the projected matrix element
\begin{equation}
t_{\sh}^{(0)}=\langle p_{j+1}|H_t|p_j\rangle,
\end{equation}
whose expansion in the doublon/bond basis is
\begin{align}
t_{\sh}^{(0)}
={}&c_\alpha^2\langle D_{j+1}|H_t|D_j\rangle
+s_\alpha^2\langle B_{j+1,\alpha}|H_t|B_{j,\alpha}\rangle \nonumber\\
&+c_\alpha s_\alpha e^{-i\phi_\alpha}
\langle D_{j+1}|H_t|B_{j,\alpha}\rangle \nonumber\\
&+c_\alpha s_\alpha e^{i\phi_\alpha}
\langle B_{j+1,\alpha}|H_t|D_j\rangle .
\end{align}
Introducing
\begin{equation}
\tau_{DB}\equiv \frac{1}{2}\left[
 e^{-i\phi_\alpha}\langle D_{j+1}|H_t|B_{j,\alpha}\rangle
+e^{i\phi_\alpha}\langle B_{j+1,\alpha}|H_t|D_j\rangle
\right],
\end{equation}
one arrives at
\begin{equation}
\begin{aligned}
t_{\sh}^{(0)}={}&c_\alpha^2\langle D_{j+1}|H_t|D_j\rangle
+s_\alpha^2\langle B_{j+1,\alpha}|H_t|B_{j,\alpha}\rangle\\
&+2\sqrt{Z_{D,\alpha}Z_{B,\alpha}}\,\tau_{DB}.
\end{aligned}
\end{equation}
For the present nearest-neighbor single-particle hopping $H_t$, the first two matrix elements vanish exactly,
\begin{equation}
\langle D_{j+1}|H_t|D_j\rangle=0,
\qquad
\langle B_{j+1,\alpha}|H_t|B_{j,\alpha}\rangle=0.
\end{equation}
One action of $H_t$ moves only one particle, so it converts a doublon into a nearest-neighbor bond state or a nearest-neighbor bond state into a neighboring doublon, but it cannot translate an entire doublon or an entire bond rigidly by one unit cell. The zeroth-order shell transfer therefore reduces to the mixed conversion contribution,
\begin{equation}
t_{\sh}^{(0)}=2\sqrt{Z_{D,\alpha}Z_{B,\alpha}}\,\tau_{DB}.
\end{equation}
This is the microscopic reason the shell transfer is nonzero in the minimal model. Transport is carried not by direct doublon-to-doublon or bond-to-bond motion, but by the hybrid character of the shell itself. More general hopping structures, or higher-order effective terms generated after additional elimination steps, could produce nonzero direct doublon or bond transfer pieces. They are absent in the present construction. This is also why $t_{\sh}^{(0)}$ is the most informative coefficient to display explicitly: among the three zeroth-order coefficients, it is the one that feeds most directly into the finite-density transport scale discussed below.

The nearest-neighbor shell interaction is defined by
\begin{equation}
U_{\sh}^{(0)}=\langle p_j p_{j+1}|H_U+H_V|p_j p_{j+1}\rangle-2\mu_{\sh}^{(0)},
\end{equation}
where $|p_jp_{j+1}\rangle$ is understood as the normalized two-shell configuration in the retained finite-density manifold.  This coefficient decomposes into projected doublon, bond, and mixed contributions in the same way. We keep $U_{\sh}^{(0)}$ at this structural level because, unlike $t_{\sh}^{(0)}$, its detailed closed form depends more strongly on which shell-pair configurations are retained in the chosen finite-density manifold. At the most elementary level one may think of
\begin{equation}
\langle DD|H_U+H_V|DD\rangle\sim U,\qquad
\langle BB|H_U+H_V|BB\rangle\sim V,
\end{equation}
while the mixed matrix elements interpolate between these limits according to the shell composition set by $c_\alpha$ and $s_\alpha$. In this sense $U_{\sh}^{(0)}$ is not a new phenomenological interaction. It is the original pair interaction $U,V$ seen through the shell projector.

\subsection{Hamiltonian Schur map and density dressing}
The density-dressed shell Hamiltonian is generated by the Hilbert-space Schur map
\begin{equation}
H_{\sh}^{(N)}\simeq P_N H P_N-P_N H Q_N(Q_N H Q_N-E_*)^{-1}Q_N H P_N.
\end{equation}
Here $E_*$ is the reference energy in the retained $N$-shell manifold.  At low shell density $\nu$ one may organize the dressing as
\begin{align}
\mu_{\sh}(\nu)&\simeq \mu_{\sh}^{(0)}+2U_{\sh}^{(0)}\nu+\delta\mu_{\rm virt}(\nu),
\\
t_{\sh}(\nu)&\simeq t_{\sh}^{(0)}+\delta t_{\rm virt}(\nu),
\\
U_{\sh}(\nu)&\simeq U_{\sh}^{(0)}+\delta U_{\rm virt}(\nu).
\end{align}
The first correction to $\mu_{\sh}$ is the Hartree shift generated by the shell interaction. The corrections $\delta t_{\rm virt}$ and $\delta U_{\rm virt}$ originate from the resolvent in the Schur self-energy
\begin{equation}
\Sigma_N(E_*)=-P_N H Q_N(Q_N H Q_N-E_*)^{-1}Q_N H P_N.
\end{equation}
Within the retained shell manifold the diagonal part of $\Sigma_N$ shifts the shell onsite energy and interaction, whereas the off-diagonal part renormalizes coherent transfer. Because the finite-density transport scale is controlled most directly by the coherent shell hopping, it is useful to write one representative off-diagonal correction explicitly,
\begin{equation}
\delta t_{\rm virt}(\nu)=-\sum_{m\in Q_N}\frac{\langle p_{j+1};\nu|H_t|m\rangle\langle m|H_t|p_j;\nu\rangle}{E_m-E_*}.
\end{equation}
This formula is worth unpacking carefully. The bra and ket states $|p_j;\nu\rangle$ and $|p_{j+1};\nu\rangle$ already include the retained-shell background at density $\nu$. The intermediate states $|m\rangle\in Q_N$ are shell-breaking configurations, for example states in which one local shell orbital is replaced by the far-detuned branch, by a higher-energy doublon/bond configuration, or by a configuration forbidden in the retained manifold but accessible virtually. The numerator therefore counts a two-step process: $H_t$ first takes the system out of the retained shell, and a second action of $H_t$ returns it to the retained shell one site away. The denominator $E_m-E_*$ is the energetic price of that excursion. The overall minus sign is the standard Schur-complement sign.

The diagonal terms may be organized schematically as
\begin{align}
\delta\mu_{\rm virt}(\nu) &\sim -\sum_{m\in Q_N}\frac{|\langle m|H_t|p_j;\nu\rangle|^2}{E_m-E_*},
\\
\delta U_{\rm virt}(\nu) &\sim -\sum_{m\in Q_N}\frac{|\langle m|H_t|p_jp_{j+1};\nu\rangle|^2}{E_m-E_*}+\cdots.
\end{align}
These corrections have a parallel meaning. The quantity $\delta\mu_{\rm virt}$ is the self-energy of one retained shell particle in the finite-density background, while $\delta U_{\rm virt}$ is the interaction renormalization associated with virtual excursions of a neighboring shell pair. The background density therefore matters twice: it changes which intermediate states are allowed, and it changes the matrix elements of the excursions that remain allowed. This is the precise sense in which filling dresses the shell instead of replacing it. Thus the finite-density shell Hamiltonian remains anchored in the same microscopic Hamiltonian: the local resonance fixes the orbital, direct projection gives the bare shell coefficients, and virtual excursions into $Q_N$ provide the density dressing.

The dissipative constraint used to isolate the finite-density slow sector is a number-conserving phase-locking jump,
\begin{equation}
J_{\sh,j}\simeq \sqrt{\kappa_\phi}
\left(p_j^\dagger+e^{-iQ}p_{j+1}^\dagger\right)
\left(p_j-e^{iQ}p_{j+1}\right).
\label{eq:Jlock_main}
\end{equation}
This jump removes the wrong relative phase without transporting shell particles.  Phase mismatch is therefore bright and fast, whereas defects inside the locked manifold are slow.

\subsection{Microscopic implementation of the dissipative channels}
\subsubsection{Boundary doublon-selective loss from a lossy auxiliary channel}\label{subsec:boundary-loss-implementation}
A natural route to boundary doublon loss is to couple the boundary doublon to a lossy auxiliary level. Let $m_1$ denote an auxiliary molecular or excited level at the boundary and let $b_q$ denote bath modes. Consider
\begin{equation}
\begin{aligned}
H_{SB}^{(d)}&=\sum_q g_q b_q^\dagger m_1+\text{H.c.},\\
H_{\rm conv}^{(d)}&=\lambda_d
(m_1^\dagger c_{1\downarrow}c_{1\uparrow}+\text{H.c.}).
\end{aligned}
\end{equation}
If the auxiliary level decays rapidly into the bath, standard Born--Markov and adiabatic-elimination steps yield an effective boundary jump
\begin{equation}
L^{(d)}=\sqrt{\gamma_d}\,c_{1\downarrow}c_{1\uparrow},
\qquad
\gamma_d\sim\frac{4\lambda_d^2}{\kappa_m},
\end{equation}
where $\kappa_m$ is the linewidth of the auxiliary channel. Projecting this jump into the shell gives
\begin{equation}
P L^{(d)}P\simeq \sqrt{\gamma_d}\,\langle D_1|p_1\rangle p_1,
\qquad
|\langle D_1|p_1\rangle|^2=Z_{D,\alpha},
\end{equation}
which is the shell jump used in the dilute branch.  The phase of $\langle D_1|p_1\rangle$ can be absorbed into the local definition of the projected jump.

\subsubsection{Number-conserving phase locking from a lossy conversion channel}
The phase-locking jump is more structured because it is number conserving inside the shell. We define $b_j=(p_j+e^{iQ}p_{j+1})/\sqrt2$ and $a_j=(p_j-e^{iQ}p_{j+1})/\sqrt2$ as the locked and mismatch shell combinations. The simplest microscopic route is then a lossy conversion from the mismatch shell combination into the locked one. Introduce a damped auxiliary bosonic mode $r_j$ and consider
\begin{equation}
H_{\rm int}^{(\phi)}=g_\phi\sum_j(r_j^\dagger b_j^\dagger a_j+a_j^\dagger b_j r_j),
\qquad
\kappa_r\sum_j \mathcal D[r_j].
\end{equation}
When $r_j$ is strongly damped, adiabatic elimination yields the effective shell jump
\begin{equation}
J_{\sh,j}=\sqrt{\kappa_\phi}\,b_j^\dagger a_j,
\qquad
\kappa_\phi\sim\frac{4g_\phi^2}{\kappa_r}.
\end{equation}
One way to see this is to use the effective-operator formalism of adiabatic elimination: the lossy auxiliary mode is the fast sector. The first virtual step converts the mismatch mode $a_j$ into an excitation of the auxiliary mode $r_j$, and the second returns that excitation to the shell as the locked mode $b_j$. This two-step process produces a dissipative conversion operator proportional to $b_j^\dagger a_j$. The crucial physical property is that $b_j^\dagger a_j$ conserves shell particle number while irreversibly converting the mismatch mode into the locked one. This microscopic route is not unique, but it makes explicit that the shell jump can be obtained as a controlled reservoir-engineered constraint rather than assumed as an ad hoc operator.

The same microscopic route also explains why the bright mismatch block acquires a gap. In the one-particle shell sector, the dissipator $\mathcal D[J_{\sh,j}]$ annihilates the locked combination but relaxes the bright combination. In other words,
\[
J_{\sh,j}|b\rangle=0,\qquad J_{\sh,j}|a\rangle\propto |b\rangle,
\]
so the mismatch channel carries a finite decay rate whereas the locked channel does not. This is the microscopic origin of the bright-block gap scale
\begin{equation}
\Gamma_{\rm lock}\sim 4\kappa_\phi,
\end{equation}
up to shell normalization conventions. It is precisely this gap that later appears in the denominator of the defect Schur return.

\section{Bright-block elimination and defect diffusion}
\label{sec:defect}
The shell Liouvillian at finite density is
\begin{equation}
\Liou_{\sh}^{(N)}[\rho] = -i[H_{\sh}^{(N)},\rho]+\sum_j \mathcal D[J_{\sh,j}]\rho + \cdots.
\end{equation}
At this stage the long-time problem is not yet purely defect-like, because shell trajectories can still visit the bright mismatch block. We denote by $V_b$ the part of the finite-density shell dynamics that couples the locked defect subspace to the bright mismatch block. The defect sector appears only after reusing the same Liouvillian Schur return, now with $P=P_{\defect}$ and $Q=Q_b$:
\begin{equation}
\begin{aligned}
\Liou_{\rm eff}^{(\defect)}\simeq{}&
P_{\defect}\Liou_{\sh}^{(N)}P_{\defect}\\
&-P_{\defect}V_bQ_b
(Q_b\Liou_{\sh,0}^{(N)}Q_b)^{-1}
Q_bV_bP_{\defect}.
\end{aligned}
\end{equation}
This is structurally identical to the dilute return, but the meaning of the fast block is now different. In the dilute problem the fast block contains the lossy states outside the edge subspace. Here $Q_b$ contains configurations with one bright mismatch mode. The fast block has changed, but the projection logic has not: the same Schur backbone is reused, while the virtual excursion changes from an edge-to-edge process to a defect-to-bright-to-defect process.

A useful way to picture one defect step is the following. Start from a locked-manifold defect configuration $|\eta\rangle$. The perturbation $V_b$ creates a bright mismatch excitation, taking the system to an intermediate state $|\mu\rangle\in Q_b$. Because this intermediate state lies in the bright block, it decays on the locking scale $\Gamma_{\rm lock}$ and may also be detuned by $\Delta_b$. A second action of $V_b$ returns the system to a new locked-manifold defect configuration $|\eta'\rangle$. It is useful to distinguish the bare defect-to-bright conversion matrix element from the effective defect hop rate generated after elimination. Let
\begin{equation}
T_{\eta\mu}^{R,L}=\langle \mu|V_b|\eta\rangle
\end{equation}
denote the conversion matrix element from a locked-manifold defect configuration to a bright intermediate state. The corresponding Schur return amplitude is
\begin{equation}
\mathcal A_{\eta'\eta}^{R,L}
=
-\sum_{\mu\in Q_b}
\frac{\langle \eta'|V_b|\mu\rangle\langle \mu|V_b|\eta\rangle}{\Gamma_\mu/2-i\Delta_\mu}.
\end{equation}
For the slow population dynamics, the corresponding nearest-neighbor defect hop rates scale as
\begin{equation}
W_{R,L}\sim
\sum_{\mu\in Q_b}
\frac{|T_{\eta\mu}^{R,L}|^2\Gamma_\mu}{(\Gamma_\mu/2)^2+\Delta_\mu^2}.
\end{equation}
A complementary minimal check of this second Schur return is given in Appendix~\ref{app:rate-benchmark}. There the locked manifold and one representative bright block
are kept explicitly, and the exact slow gap of that unreduced block is
compared directly with the Schur estimate derived here.

If one dominant bright block controls the return, this reduces to
\begin{equation}
W_{R,L}\simeq \frac{|T_{\rm db}^{R,L}|^2\Gamma_{\rm lock}}{(\Gamma_{\rm lock}/2)^2+\Delta_b^2},
\label{eq:SM_Wdef}
\end{equation}
which is the reduced rate formula used to define the defect diffusion constant below. The denominator is physically transparent: the real part $\Gamma_{\rm lock}/2$ is the dissipative cost of visiting the bright block, while the imaginary part $\Delta_b$ is the coherent detuning of that intermediate state. This is why defect motion is intrinsically a second-order process. A defect can move only by briefly borrowing a bright mismatch and then returning to the locked manifold. In a slightly more general notation, allowing the dominant bright block on the two sides to have different detunings $\Delta_{b,R}$ and $\Delta_{b,L}$, one obtains
\begin{equation}
\frac{W_R}{W_L}\simeq
\frac{|T_{\rm db}^{R}|^2}{|T_{\rm db}^{L}|^2}
\frac{(\Gamma_{\rm lock}/2)^2+\Delta_{b,L}^2}{(\Gamma_{\rm lock}/2)^2+\Delta_{b,R}^2}.
\label{eq:SM_WRL_ratio}
\end{equation}
This expression makes explicit that the defect walk becomes nonreciprocal either because the defect couples with different strength to the right and left bright blocks, or because the corresponding bright intermediate states are detuned differently.

At the symmetric working point,
\begin{equation}
T_{\rm db}^R=T_{\rm db}^L\equiv T_{\rm db},
\qquad
W_R=W_L\equiv W_{\defect},
\end{equation}
so the effective defect generator reduces to a symmetric nearest-neighbor random walk. On the lattice this is the discrete Laplacian acting on the defect probability amplitude or, at the level of slow expectation values, on the corresponding defect density field. The diffusion constant is then
\begin{equation}
D_{\defect}\equiv W_{\defect}\simeq \frac{|T_{\rm db}|^2\Gamma_{\rm lock}}{(\Gamma_{\rm lock}/2)^2+\Delta_b^2}.
\label{eq:SM_Ddef_sharp}
\end{equation}
If the dominant defect-to-bright conversion matrix element is set by the shell hopping, then $T_{\rm db}\sim t_{\sh}(\nu)$ and one obtains the resonant-locking estimate
\begin{equation}
D_{\defect}\simeq \frac{4|t_{\sh}(\nu)|^2}{\Gamma_{\rm lock}},
\qquad |\Delta_b|\ll \Gamma_{\rm lock}.
\label{eq:SM_Ddef_resonant}
\end{equation}
When $W_R\neq W_L$, the same Schur return produces a nonreciprocal defect walk. We defer its spectrum, skin profile, and finite-size offset to Sec.~\ref{sec:skin}, and first isolate the symmetric point $W_R=W_L$ where the pure diffusive law is exposed.

The reduction is controlled only when the bright-block gap exceeds both the defect-transfer scale and the leading shell-breaking corrections. In that regime the slow finite-density object is not phase mismatch itself, because the latter is bright and fast, but a defect internal to the locked manifold.

Once the defect generator has reduced to a symmetric walk, the finite-size scaling follows in the standard way. For an open chain of length $L$, the eigenmodes of the discrete Laplacian are
\begin{equation}
\varphi_n(j)=\sin\frac{n\pi j}{L+1},\qquad n=1,2,\ldots,L,
\end{equation}
with eigenvalues
\begin{equation}
\lambda_n\simeq -4D_{\defect}\sin^2\frac{n\pi}{2(L+1)}.
\end{equation}
The slowest nonzero mode is $n=1$, so for large $L$ one obtains
\begin{equation}
\Delta_L\equiv -\lambda_1\simeq 4D_{\defect}\sin^2\frac{\pi}{2(L+1)}
\simeq \frac{\pi^2D_{\defect}}{(L+1)^2}.
\end{equation}
This is the diffusive finite-size law used above.  The same reduced defect mode also gives linear mean-square spreading in the time domain; this complementary diagnostic is given in Appendix~\ref{app:defect-msd}.  The logic is parallel to the dilute case, but the slow object and the eliminated block have changed.  In the dilute branch the slow pole is controlled by edge-to-edge communication through a lossy boundary sector; in the finite-density branch it is controlled by defect motion through virtual excursions into the bright mismatch block.

\subsection{Microscopic shell-many-body bridge}
\label{subsec:manybody-bridge}
\begin{figure}[!t]
\centering
\safeincludegraphics[width=0.5\textwidth]{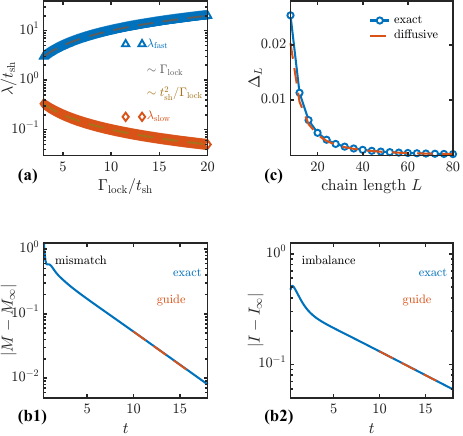}
\caption{Finite-density diagnostics at the symmetric locking point.  The upper-left panel (a) shows the bright-defect scale separation, with $\lambda_{\rm fast}\propto\Gamma_{\rm lock}$ and $\lambda_{\rm slow}\propto t_{\sh}^2/\Gamma_{\rm lock}$.  The upper-right panel (c) compares the exact open defect-sector gap with the diffusive law $\pi^2D_{\defect}/(L+1)^2$.  The lower panels show the unreduced shell-many-body traces: (b1) the connected mismatch $|M(t)-M_\infty|$ and (b2) the connected imbalance $|I(t)-I_\infty|$.  The late-time guide uses $\lambda_{\rm guide}=\Delta_L^{(\defect)}$ at the same parameters.}
\label{fig:fig3}
\end{figure}

The reduced defect generator is the cleanest object from which the diffusive $L^{-2}$ law follows, but it is also useful to verify directly that the unreduced shell-many-body dynamics already exhibits the same fast-slow hierarchy. The representative microscopic traces are shown in Figs.~\ref{fig:fig3}(b1) and~\ref{fig:fig3}(b2). We define the corresponding observables here so that the finite-density diagnostics are self-contained.

We evolve the full finite-density shell Liouvillian $\Liou_{\sh}^{(N)}$ in a representative fixed-$N$ sector and monitor two connected observables. The first is the connected mismatch
\begin{equation}
|M(t)-M_\infty|,
\qquad
M(t)=\sum_j \langle J_{\sh,j}^\dagger J_{\sh,j}\rangle_t,
\end{equation}
which measures how rapidly bright phase mismatch is removed. The second is the connected half-chain imbalance
\begin{equation}
|I(t)-I_\infty|,
\qquad
I(t)=\sum_{j\le L/2}\langle n_j\rangle_t-\sum_{j>L/2}\langle n_j\rangle_t,
\end{equation}
which probes the slower redistribution of shell density after the bright sector has relaxed. Here $M_\infty$ and $I_\infty$ are taken from the exact steady state of the same finite-$N$ Liouvillian.

In Figs.~\ref{fig:fig3}(b1) and~\ref{fig:fig3}(b2), the lower-left panel displays the rapid decay of the connected mismatch, while the lower-right panel displays the slower connected imbalance tail in the same microscopic shell-many-body evolution. In both subpanels the late-time guide is drawn with the reduced defect-sector rate $\lambda_{\rm guide}\equiv\Delta_L^{(\defect)}$ computed at the same parameters from the effective generator of Sec.~\ref{sec:defect}. These traces show that the slow tail visible before the final projection already lies on the same rate scale as the projected defect mode. They serve as a microscopic bridge to the reduced theory, while the clean $L^{-2}$ law follows from the isolated defect generator obtained after bright-block elimination.

\section{Asymmetric defect-sector skin walk}
\label{sec:skin}

Away from symmetric locking, $W_R\ne W_L$ and the projected defect generator becomes an asymmetric walk.  Its bulk dispersion is
\begin{equation}
\lambda(q)=-(W_R+W_L)+W_Re^{-iq}+W_Le^{iq}.
\label{eq:asym_disp_main}
\end{equation}
For an open chain, a similarity transformation makes the hopping reciprocal and gives eigenvalues
\begin{equation}
\lambda_n^{\rm asym}=-(W_R+W_L)+2\sqrt{W_RW_L}
\cos\frac{n\pi}{L+1}.
\label{eq:asym_eigs_main}
\end{equation}
The right eigenmodes have the open-chain skin profile
\begin{equation}
\psi_{n,j}^{(R)}\propto
\left(\sqrt{\frac{W_R}{W_L}}\right)^j
\sin\frac{n\pi j}{L+1},
\qquad
\xi_{\rm skin}^{-1}=\frac12\left|\ln\frac{W_R}{W_L}\right|.
\label{eq:skin_main}
\end{equation}
Equivalently, the open-chain slow scale contains the finite nonreciprocal offset
\begin{equation}
\Delta_L^{\rm asym}=\left(\sqrt{W_R}-\sqrt{W_L}\right)^2
+4\sqrt{W_RW_L}\sin^2\frac{\pi}{2(L+1)}.
\label{eq:asym_gap_main}
\end{equation}
The symmetric locking point studied in Sec.~\ref{sec:defect} is therefore the reciprocal point of this defect-sector skin walk, where the offset and the nonreciprocal skin factor vanish and the pure diffusive $L^{-2}$ law remains.  This skin walk is not imposed at the level of the original hopping Hamiltonian.  It is generated after the finite-density bright block is eliminated.  A minimal benchmark is given in Appendix~\ref{app:skin-benchmark}, and the physical interpretation follows the standard asymmetric-hopping skin mechanism \cite{HatanoNelson1996,YaoWang2018,Kunst2018}.

\section{Discussion and summary}
\label{sec:discussion}

The main result is a microscopic selection principle for Liouvillian slow sectors. The effective SSH channel is one consequence of projecting the microscopic hopping onto the resonant shell, but the broader structure is the joint projection of the Hamiltonian and reservoir algebra onto the same parent object. Once the local doublon-bond shell has been identified, the same projection backbone explains why boundary loss, shell criticality, and finite-density phase locking lead to different slow sectors.  The boundary-loss protocol removes a lossy sector and leaves an edge-memory pole; at shell criticality the topological edge hierarchy collapses and a near-zero coherent doublet is exposed; the phase-locking protocol removes bright mismatch and leaves defects as the asymptotic degrees of freedom.  These are not separate effective models, but different projected returns of one microscopic resonant shell.

Some ingredients are model specific, such as the particular branch splitting generated by $H_\delta$ and $H_h$ and the spin-dependent hopping pattern used to obtain a simple branch-selective channel.  The logic is more general.  Whenever an interacting local resonance produces a composite orbital with both reservoir visibility and mobility, reservoir engineering can select distinct slow Liouvillian sectors by changing the eliminated fast block.  The Schur return itself is standard; the physical content lies in identifying the microscopic shell and in tracking how the Hamiltonian and jump projections act on that same object.

The same reductions also specify the regime of validity. The shell projection requires a branch separation large enough to isolate one local resonant composite.  The dilute edge-memory law assumes the Zeno window $|\varepsilon_{LR}|\ll\Gamma_D$.  The shell-critical spacing is the coherent spacing of the selected near-zero doublet, not the full Liouvillian gap of the boundary-damped ladder.  The finite-density diffusion law assumes that the bright mismatch block relaxes faster than the defect-transfer scale and the leading shell-breaking corrections.  Outside these windows one should keep more shells or more bright blocks explicitly.

The mechanism can be tested at three levels.  The two-particle spectrum should show branch-selective dimerization generated by the microscopic spin-dependent hopping and bond splitting.  In the dilute regime, boundary doublon loss should produce a slow edge-memory pole controlled by the edge-to-edge shell overlap and the doublon visibility.  At finite shell filling, phase locking should produce fast mismatch decay followed by a slower imbalance tail governed by the defect-sector rate and by the $L^{-2}$ finite-size gap.  The nonreciprocal extension predicts a further crossover from pure defect diffusion to a skin-offset-dominated slow scale when $W_R\neq W_L$.  Together these signatures provide direct tests of the shell-first selection principle in open correlated lattices.

\begin{acknowledgments}
We acknowledge support from the National Natural Science Foundation of China under Grants No.~12275193 and No.~11975166.
\end{acknowledgments}

\section*{Data availability}
The data that support the findings of this article are available from the author upon reasonable request.

\FloatBarrier
\appendix

\section{Reduced-shell real-time diagnostic of defect diffusion}
\label{app:defect-msd}
The main text extracts the asymptotic law from the defect-sector gap because that is the cleanest route to the exponent. For the finite-density branch it is nevertheless useful to display one complementary time-domain diagnostic inside the reduced defect description. We therefore plot the mean-square displacement
\begin{equation}
\mathrm{MSD}(t)=\sum_j (j-j_0)^2 n_j(t),
\end{equation}
of a defect packet initially centered at $j_0$ in the symmetric locked branch, where $n_j(t)$ denotes the normalized defect-density profile in the reduced description. For diffusive motion one expects
\begin{equation}
\mathrm{MSD}(t)\simeq 2D_{\defect}\,t,
\end{equation}
with the same diffusion constant $D_{\defect}$ that governs the open-chain gap in Sec.~\ref{sec:defect}.

This reduced-shell diagnostic is not used as the primary extraction of the asymptotic exponent. Its role is more limited and more concrete: it shows directly in the time domain that the same reduced defect mode responsible for the $L^{-2}$ finite-size law also produces linear-in-time defect spreading.
\FloatBarrier

\section{Minimal locked-plus-bright rate benchmark}
\label{app:rate-benchmark}
\begin{figure}[!t]
\centering
\safeincludegraphics[width=0.8\columnwidth]{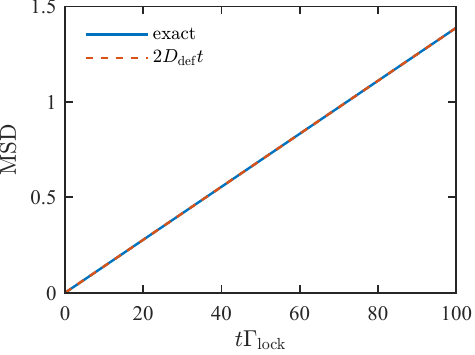}
\caption{Reduced-shell real-time diagnostic of defect diffusion. The plotted quantity is the mean-square displacement $\mathrm{MSD}(t)=\sum_j (j-j_0)^2 n_j(t)$ of a defect packet initially centered at $j_0$ in the symmetric locked branch. The exact reduced-shell evolution is compared with the diffusive law $\mathrm{MSD}\simeq 2D_{\defect}\,t$, using the same diffusion constant $D_{\defect}$ that enters the finite-size gap law derived in Sec.~\ref{sec:defect} and used above.}
\label{fig:s2}
\end{figure}
The finite-density argument in Sec.~\ref{sec:defect} uses the projected defect
generator because that is the cleanest object from which the asymptotic
$L^{-2}$ law follows. It is nevertheless useful to retain one minimal
unreduced block that still contains both the locked manifold and the bright
intermediate sector, so that the second Schur return can be checked before
passing all the way to the reduced defect walk. We therefore consider a
minimal locked-plus-bright benchmark in which the slow sector is represented
by a defect chain of length $L$ and the fast sector by one representative
bright block characterized by the same $T_{\rm db}$, $\Gamma_{\rm lock}$, and
$\Delta_b$ that enter Eq.~\eqref{eq:SM_Wdef}. The exact slow gap of this
unreduced block is obtained by direct diagonalization, while the Schur
prediction is evaluated from
\begin{equation}
\begin{aligned}
W_{\defect}&=\frac{|T_{\rm db}|^2\Gamma_{\rm lock}}{(\Gamma_{\rm lock}/2)^2+\Delta_b^2},\\
\Delta_{\rm Schur}&=4W_{\defect}\sin^2\frac{\pi}{2(L+1)}.
\end{aligned}
\end{equation}

\begin{figure*}[!t]
\centering
\safeincludegraphics[width=0.82\textwidth]{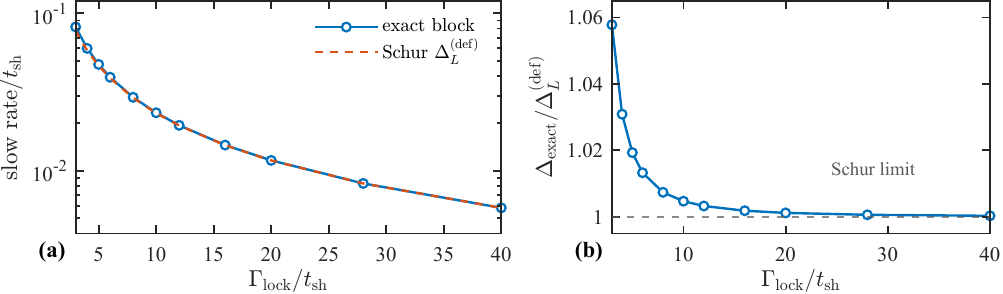}
\caption{Minimal locked-plus-bright rate benchmark for the second Schur
return. The benchmark keeps the slow locked manifold together with one
representative bright intermediate block and compares the exact slow gap of
that unreduced problem with the Schur estimate. The data shown here use
$L=12$, $t_{\sh}=1$, $T_{\rm db}=1$, and $\Delta_b=0$. Panel (a) compares the exact slow gap with the Schur estimate as a function of
$\Gamma_{\rm lock}/t_{\sh}$, while panel (b) shows the ratio between the two. Once $\Gamma_{\rm lock}/t_{\sh}$ is moderately large, the exact
result rapidly approaches the projected-rate prediction, validating the use
of Eq.~\eqref{eq:SM_Wdef} as the finite-density slow scale.}
\label{fig:s3}
\end{figure*}

The resulting comparison is shown in Fig.~\ref{fig:s3}. At small
$\Gamma_{\rm lock}/t_{\sh}$ the elimination is only moderately controlled, so
the exact slow gap lies slightly above the Schur estimate. As the locking
scale grows, the ratio rapidly approaches unity. This is the behavior
expected for a controlled second return. The benchmark is therefore not a
replacement for the many-body shell numerics used above. Rather, it provides an intermediate check between the direct microscopic bridge of Sec.~\ref{sec:defect}\,A and the fully reduced defect generator of Sec.~\ref{sec:defect}. It shows that the projected rate formula already captures the slow scale accurately once
the bright block is sufficiently fast, and it makes explicit the control
criterion implicitly assumed in the main text, namely that the bright-block
relaxation scale must exceed both the defect-transfer scale and the leading
shell-breaking corrections.

\section{Minimal asymmetric defect-skin benchmark}
\label{app:skin-benchmark}
\begin{figure*}[!t]
\centering
\safeincludegraphics[width=0.82\textwidth]{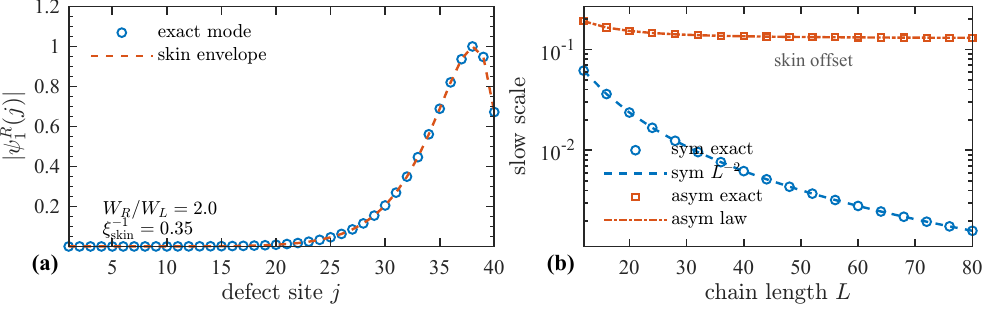}
\caption{Minimal open-chain benchmark for the asymmetric defect-sector skin walk. Panel (a) compares the exact right eigenmode of the slowest open-chain mode with the analytic open-chain skin-mode profile $(\sqrt{W_R/W_L})^j\sin[\pi j/(L+1)]$ for $W_R/W_L=2$. Panel (b) compares the finite-size slow scale of a reciprocal walk with that of an asymmetric walk. The reciprocal case follows the $L^{-2}$ diffusive law, whereas the asymmetric case approaches the finite skin offset $(\sqrt{W_R}-\sqrt{W_L})^2$.}
\label{fig:s4}
\end{figure*}
The asymmetric defect generator introduced by Eq.~\eqref{eq:SM_WRL_ratio} and analyzed in Sec.~\ref{sec:skin} has a direct open-chain consequence: its right and left eigenvectors acquire opposite exponential skin factors. To isolate this effect from all other microscopic details, we consider the reduced open defect walk with right and left rates $W_R$ and $W_L$.  The generator is
\begin{equation}
\begin{aligned}
K_{\rm skin}={}&\sum_{j=1}^{L-1}\left(W_R|j+1\rangle\langle j|+W_L|j\rangle\langle j+1|\right)\\
&-(W_R+W_L)\sum_{j=1}^{L}|j\rangle\langle j| .
\end{aligned}
\end{equation}
For $W_R\neq W_L$, the spectrum is still given by Eq.~\eqref{eq:asym_eigs_main}, and the right eigenvectors are skin-localized according to the envelope derived in Sec.~\ref{sec:skin}. The slow scale can also be written as
\begin{equation}
\Delta_L^{\rm asym}=(\sqrt{W_R}-\sqrt{W_L})^2+
4\sqrt{W_RW_L}\sin^2\frac{\pi}{2(L+1)} .
\end{equation}
The first term is the nonreciprocal offset, while the second is the residual finite-size diffusion scale. At the reciprocal point $W_R=W_L$ the offset vanishes and the $L^{-2}$ law is recovered.

Figure~\ref{fig:s4} provides the corresponding minimal numerical check. It is not a new microscopic assumption. Rather, it is the open-chain realization of the same biased defect generator already obtained from the second Schur return. It shows explicitly that the symmetric locking point used in the main text is the reciprocal point of an underlying defect-sector skin walk, so the skin result is a controlled specialist consequence of the same defect return rather than an additional assumption.

\FloatBarrier
\clearpage
\bibliography{PRL_submission_refs_final}

\end{document}